\documentclass[12pt]{article}

\def\d{\displaystyle}
\def\gapprox{\:\raisebox{-1.25ex}{$\stackrel{\textstyle>}{\approx}$}\:}

\begin{document}
\noindent Dec 2012 \hfill ICTS-TIFR \ 2012/13\\
\hspace*{9.7 cm}          TIFR/TH/12-47\\
\bigskip

\begin{center}
{\Large{\bf A Study of $U(N)$ Lattice Gauge Theory in
2-dimensions}} \\[1cm]
Spenta R. Wadia \\
International Centre for Theoretical Sciences (ICTS-TIFR)\\
Tata Institute of Fundamental Research\\
TIFR Centre Bldg, IISc Campus \\
Bangalore 560012 India\\
and\\
Dept of Theoretical Physics\\
Tata Institute of Fundamental Research\\
Homi Bhabha Rd, Mumbai 500004 India
\end{center}
\bigskip

\begin{center}
{\Large{\bf Abstract}}
\end{center}
\bigskip
This is an edited version of an unpublished 1979 EFI (U. Chicago) preprint

\newpage

\noindent July, 1979 \hfill EFI \ 79/44
\bigskip

\begin{center}
{\Large{\bf A Study of $U(N)$ Lattice Gauge Theory in
2-dimensions$^\star$}} \\[1cm]
Spenta Wadia \\
The Enrico Fermi Institute \\
The University of Chicago \\
Chicago, Illinois, 60637
\end{center}
\bigskip

\begin{center}
{\Large{\bf Abstract}}
\end{center}
\bigskip

The $U(N)$ lattice gauge theory in 2-dimensions can be considered as
the statistical mechanics of a Coulomb gas on a circle in a constant
electric field.  The large $N$ limit of this system is discussed and
compared with exact answers for finite $N$.  Near the fixed points of
the renormalization group and especially in the critical region where
one can define a continuum theory, computations in the thermodynamic
limit $(N \rightarrow \infty)$ are in remarkable agreement with those
for finite and small $N$.  However, in the intermediate coupling
region the thermodynamic computation, unlike the one for finite $N$,
shows a continuous phase transition. 
This transition seems to be a pathology of the infinite $N$ limit and
in this simple model has no bearing on the physical continuum limit.

\vfill

\noindent $^\star$Work supported in part by the NSF: Grant
No. PHY-78-01224. 

\newpage

\noindent {\bf Introduction}
\bigskip

The $1/N$ expansion has proved very useful in studying the physical
spectrum of $N$ component spin systems ($\sigma$-models) and several
other tractable field theories with large dimension of internal
symmetry group, in two dim.$^1$ It is used on
the assumption that the qualitative, and, in certain cases, even
quantitative character of the spectrum is $N$ independent.  Then
taking the large $N$ limit enormously simplifies the computation since
one can use a combination of a steepest descent and mean field
approach in which each component of the spin interacts with the mean
field of the other components.

For $SU(N)$ color gauge theories of the strong interaction, the $1/N$
expansion was introduced by `t Hooft$^2$ who showed
that in the Feynman graph expansion of the theory to each order in
$g^2N$ (fixed as $N \rightarrow \infty$), the leading contribution
comes from planar diagrams.  Diagrams with holes (quark loops) and
handles are suppressed by factors of $1/N$ and $1/N^2$ respectively.
In this sense, the expansion in $1/N$ is analogous to the topological
expansion of the dual model and as $N \rightarrow \infty$ the mesonic
and gluonic states appear as free particle states.
Witten$^3$ has recently incorporated baryons as
solitons into this scheme and has enumerated and emphasized the
qualitative agreement between simple and basic aspects of hadronic
spectroscopy and the $1/N$ expansion.  However, the computational
difficulties of planar QCD remain intractable in the diagrammatic
language of its formulation, though some remarkable progress in the
counting of planar diagrams for quartic and cubic vertices has been
made by Br\'ezin et al., using a WKB approach in
$1/N$.$^4$ 

Now in the statistical mechanics of $N$ component spin systems the
relation of the large $N$ limit to a mean field computation has
received precise mathematical formulation by the introduction of a
Lagrange multiplier as an auxiliary field.  It is desirable to have
an analogous formulation in the case of QCD, which avoids direct
encounters with the diagram technique.  Further, it is desirable to
evaluate the relevance of the large $N$ limit. The question is very
simple: to what extent and in which domains of coupling is a
thermodynamic computation in which $N \rightarrow \infty$, relevant
for the case of finite $N$.

We present in this paper a study of $U(N)$ lattice gauge theories in
2-dimensions to clarify these questions.  Our method follows studies
in the statistical theories of spectra and ref. (\ref{four}).

In two dimensions, since plaquette variables are the independent ones,
the partition function reduces to that of a random unitary matrix with
$N^2$ degrees of freedom.  This in turn is the partition function of a
Coulomb gas on a circle in a constant electric field $E = 2\beta/N$.
The problem of computing the free energy and other thermodynamic
quantities as $N \rightarrow \infty$ reduces to the solution of a
singular integral equation for the charge density, with a Cauchy
kernel.  This problem is always exactly soluble.  The solution depends
on the strength of the electric field $E$.  For $0 \leq E \leq 1$ there
is no gap in the charge distribution and the charge density is
conjugate to the electric field.  For $1 \leq E < \infty$ the charge
distribution has a gap and the problem reduces to the Riemann-Hilbert
problem for a simple arc in the plane.  The appearance of a gap in the
charge density signals a continuous phase transition at $E = 1$.  
An order parameter
that measures the randomness of the system has a constant value in the
`strong coupling' phase $0 \leq E \leq 1$ and goes to zero for large
$E$ in the spin wave phase. The string constant is computed and a
simple renormalization group used to discuss the physically relevant
continuum limit.

The Coulomb gas partition function is also a Toeplitz determinant
involving Bessel functions.  For finite $N$ it is obviously an
analytic function of the temperature $\beta$.  Taylor expansions
around the fixed points $\beta = 0$ and $\beta = \infty$ indicate an
excellent agreement with the computation for $N \rightarrow \infty$.
To us, this agreement is far from self-evident.  Near $\beta =
\infty$, this can be seen to result from a scaling argument.

An interesting mathematical corollary is that the problem of
evaluating Toeplitz determinant of order $N$, can always be mapped to
the problem of a  Coulomb gas on a circle, which in the $N = \infty$
limit is always exactly soluble.  Toeplitz determinants are
encountered in the Ising model and their evaluation is usually done by
using Szeg\"o's theorem (c.f. ref. (\ref{eleven})).  

\newpage

\noindent {\bf I. Equivalence of $U(N)$ Gauge Theory in
2-dimensions to the Statistical Mechanics of a 1-dimensional Coulomb gas}

\bigskip

The partition function of the $U(N)$ lattice gauge theory is defined
by$^5$ 
\begin{equation}
Z(V,N,\beta) = \int d\mu \exp\left[\beta \sum_\rho \left(U(P) -
N\right)\right] 
\label{one}
\end{equation}
$d\mu$ is a real measure over the gauge group elements at each link;
the summation over $P$ is over all oriented plaquettes and $U(P)$ is a
product of group elements, in the fundamental representation, at the 4
oriented links of $P$.  $U^+(P)$ would be the group element associated
with the links in the opposite direction.  $\beta = \d{1 \over g^2_0}$
is the temperature and $V$ is the volume of the system in lattice
units. 

In two space-time dimensions it is possible to treat plaquette
variables as independent.$^6$  A simple way to see this is to fix the
axial gauge in $Z$: all time-like links are chosen to be one and all
space-like links at one particular time `$t_0$' are chosen to be one.
Since the time links are trivial the space-like links at time `$t$'
can be expressed as a unique product of a time ordered string of
independent plaquette variables, the expression for $Z$ in (\ref{one}) becomes
an integral over a single unitary matrix 
\[
Z = \left[\int dU e^{\beta\left[{\rm Tr} U + {\rm Tr}
U^+\right]}\right]^V e^{-2\beta NV}
\]

Hence, the problem is reduced to the computation of the partition
function of a random unitary matrix with $N^2$ real independent
variables 
\begin{equation}
z = \int dU e^{\beta({\rm Tr}U + {\rm Tr}U^+)}
\label{two}
\end{equation}
Integration over the group $U(N)$ is known.$^7$  It is convenient to
make a separation of variables into group invariant and non-invariant
parameters.  Let $R$ be the unitary matrix that brings $U$ to diagonal
form 
\[
U = R^+ DR, \ \ \ \ \ D_{ij} = \delta_{ij} e^{i\theta_j}
\]
$R$ is arbitrary up to an invariant subgroup of $U(N)$, hence, after
appropriate choice of gauge it depends on $N^2 - N$ parameters and 
\begin{equation}
dU = C_N dR \prod^N_{i=1} {d\theta_i \over 2\pi} \prod_{i<j}
\left|e^{i\theta_i} - e^{i\theta_j}\right|^2
\label{three}
\end{equation}
Substituting (\ref{three}) in (\ref{two}) we get
\begin{equation}
z = {1 \over N!} \int^{2\pi}_0 \prod^N_{i=1} {d\theta_i \over 2\pi}
e^{N^2\left[{2\beta \over N} \d\sum^N_{i=1} {1 \over N} \cos \theta_i +
2 \d\sum_{i<j} {1 \over N^2} \log |e^{i\theta_i} -
e^{i\theta_j}|\right]} 
\label{four}
\end{equation} 
The factor $N$! appeared by choice of $C_N$ in (\ref{three}) such that
$z = 1$ for $\beta = 0$.

Now define an action
\begin{equation}
-S = {\beta \over N} \sum^N_{i=1} {1 \over N} \cos \theta_i +
\sum_{i<j} {1 \over N^2} \log \left|e^{i\theta_i} -
e^{i\theta_j}\right|
\label{five}
\end{equation}
then
\begin{equation}
z = {1 \over N!} \int^{2\pi}_0 \prod^N_{i=1} {d\theta_i \over 2\pi}
e^{-2SN^2} 
\label{six}
\end{equation}
The action $S = o(1)$ in $N$ provided $\beta/N = E/2$ is a fixed
parameter of the theory.  If so, then for large $N$ we can consider
evaluating $z$ by a WKB expansion in the small parameter $1/N^2$.

The action $S$ in (\ref{five}) is essentially the electrostatic
potential energy of $N$ charges of strength $1/N$ distributed on a
unit circle in the plane, in the presence of a constant electric field
$E/2$ along the $x$-axis.  The repulsive logarithmic interaction tends
to distribute the charge uniformly about the circle and the electric
field tends to drive all $\theta_i \rightarrow 0$.  There is another
analogy with spin systems.  Define a unit spin vector 
\[
\vec\sigma_i = (\cos \theta_i, \sin \theta_i)
\]
then 
\[
-S = {H \over 2} \sum^N_{i=1} {1 \over N} \sigma^x_i + \sum_{i<j} {1
\over N^2} \log\left(1 - \vec\sigma_i \cdot \vec\sigma_j\right)
\]
which is the potential energy of unit spins equally spaced along the
$x$-axis in a constant magnetic field $H/2 = \beta/N$.  The logarithm
represents a continuous version of an anti-ferromagnetic coupling.  In
the absence of $H$, $\langle \vec\sigma_n \rangle = \left(\cos {2\pi n
\over N}, \sin {2\pi n\over N}\right)$.  In the rest of this paper we
shall only consider the electrostatic analogy.

\newpage

\noindent {\bf II. WKB and Mean Field Calculation}
\bigskip

If we define the free energy per degree of freedom $f(E)$ by $z =
\exp(N^2 f(E))$, then we have the WKB expansion in $1/N^2$
\begin{equation}
f(E) = -2S_0 + {S_1 \over N^2} + 0\left({1 \over N^4}\right)
\label{seven}
\end{equation}
$S_0$ is the minimum value of the potential energy $S$ and $S_1$ is
the result of gaussian fluctuations around the most probable
configuration that minimizes $S$.  For the present, we shall be
concerned with evaluating $F(E)$ as $N \rightarrow \infty$.

As we have noted, the point charges on the unit circle carry a charge
$= 1/N$.  The interaction energy between two charges is $0\left(\d{1
\over N^2}\right)$, which means that they are very weakly coupled.
However, the interaction energy of any one charge in the mean field of
the other $N^2 - N -1 \approx N^2$ charges is $0(1)$.  This motivates
a mean field theory type computation$^7$ in which one introduces a
macroscopic charge density $u(\theta) \geq 0$, on the unit circle,
such that $u(\theta)d\theta/2\pi$ is the fraction of the number of
charges between $\theta$ and $\theta + d\theta$.  By definition
\begin{equation}
\int^{2\pi}_0 u(\theta) {d\theta \over 2\pi} = 1
\label{eight}
\end{equation}
In this continuum limit as $N \rightarrow \infty$ we have
\begin{equation} 
-S[u] = {E \over 2} \int^{2\pi}_0 {d\theta \over 2\pi} u(\theta) \cos
\theta + P \int^{2\pi}_0 {d\theta \over 2\pi} \int^{2\pi}_0 {d\phi
\over 2\pi} u(\theta) \log |e^{i\theta} - e^{i\phi}| u(\phi)
\label{nine}
\end{equation}
$P$ stands for the Cauchy principal value.

The most probable charge distribution is a stationary point of
(\ref{nine}) subject to the normalization (\ref{eight}).  Hence, this
distribution satisfies the following singular integral equation 
\[
{E \over 2} \cos\phi + P \int^{2\pi}_0 {d\theta \over 2\pi} u(\theta)
\log |e^{i\theta} - e^{i\phi}| = \lambda
\]
$\lambda$ is a lagrange multiplier corresponding to (\ref{eight}).  It
represents the average potential at $\phi$ due to all the other
charges, and it is $\phi$ independent.  Taking $\phi$ derivatives we
have 
\begin{equation}
E \sin \phi = - P \int^{2\pi}_0 {d\theta \over 2\pi} u(\theta) \cot
\left({\theta - \phi \over 2}\right)
\label{ten}
\end{equation}
which states that the tangential forces on the charge at $\phi$ due to
all the other charges balance the force of the electric field.
\bigskip

\noindent {\bf III. Solution of Integral Equation}
\bigskip

Let us begin by assuming that the function $u(\theta)$ is smooth in
the sense that $u'(\theta)$ is continuous for $0 < \theta < 2\pi$.
The solution to (\ref{ten}) can be immediately written down if we
recall the Hilbert transform formula for the circle:

If $\Phi(Z)$ is analytic for $|Z| \leq 1$, then on the unit circle we
have 
\[
\Phi(e^{i\theta}) = \Phi(Z=0) + {P \over 2\pi i} \int^{2\pi}_0
d\theta\Phi(e^{i\theta}) \cot\left({\theta - \phi \over 2}\right)
\]
The choice $\Phi(Z) = 1 + EZ$ gives the correctly normalized solution 
\begin{equation}
u(\theta) = 1 + E \ \cos \ \theta
\label{eleven}
\end{equation}
The condition $u(\theta) \geq 0$ for $0 \leq \theta \leq 2\pi$,
however, requires $0 \leq E \leq 1$.

When $E=0$, $u(\theta) = 1$ as expected; turning on the electric field
along the $x$-direction diminishes the density around $\theta = \pi$
and produces a crouching of charge near $\theta = 0$.  Also $u(\theta)
= u(-\theta)$.  Note that $u(\theta)$ is smooth and never vanishes for
$E < 1$.  At $E = 1$ the distribution hits a zero at the single point
$\theta = \pi$.  This gives us a hint that as $E$ starts exceeding the
critical value 1, there would start appearing a gap in the charge
distribution around $\theta = \pi$ which would increase as a
continuous and monotonic function of $E$, $1 < E < \infty$.  In the
following we shall establish this picture by solving the integral
equation (\ref{ten}) with the boundary condition that $u(\theta) = 0$
\underbar{outside} the interval $-\alpha < \theta < \alpha$.  For
$\alpha \neq \pi$ we do not assume $u(\theta)$ to be smooth at the end
points $\theta = \alpha$.  The integral equation now becomes
\begin{eqnarray}
E \sin \phi &=& - P \int^\alpha_{-\alpha} {d\theta \over 2\pi}
u(\theta) \cot \left({\theta - \phi \over 2}\right) \nonumber \\[2mm] 
&& \int^\alpha_{-\alpha} {d\theta \over 2\pi} u(\theta) = 1 
\label{twelve}
\end{eqnarray}

To solve (\ref{twelve}) we introduce complex notation $t =
e^{i\theta}$ and $t_0 = e^{i\phi}$.  (\ref{twelve}) becomes 
\begin{eqnarray}
E(t_0 - t^{-1}_0) &=& {P \over \pi i} \int^{t_2}_{t_1} {dt \over t}
{t+t_0 \over t-t_0} u(t) \nonumber \\[2mm] 
&& {1 \over 2\pi} \int^{t_2}_{t_1} {dt \over it} u(t) = 1, \ \ \ t_1 =
t^\star_2 = e^{-i\alpha} 
\label{thirteen}
\end{eqnarray}
Now the important point is that the normalization condition enables us
to write the integral equation in the form 
\[
E(t_0 - t^{-1}_0) = {P \over \pi i} \int^{t_2}_{t_1} {dt \over t-t_0}
2u(t) - 1 
\]
The kernel is Cauchy type and we have adapted the general formalism to
solve such equations to our problem.$^9$  One starts out by
considering the function 
\begin{equation}
\Phi(z) = {1 \over 2\pi i} \int^{t_2}_{t_1} {dt \over t} {t+z \over
t-z} u(t)
\label{fourteen}
\end{equation}
analytic in the plane cut by the $(t_1,t_2)$.  The direction of the
arc is assumed counterclockwise.  Denote the boundary values of this
function as $z$ approaches a point $t$ on the arc $(t \neq t_1,t_2)$,
from the left and right by $\Phi^+(t)$ and $\Phi^-(t)$ respectively.
Then by the Plemelj formulae$^8$ for the boundary values of analytic
functions we have 
\begin{equation}
\Phi^+(t) - \Phi^-(t) = 2u(t)
\label{fifteen}
\end{equation}
\begin{equation}
\Phi^+(t) + \Phi^-(t) = {P \over \pi i} \int^{t_2}_{t_1} {d\xi \over
\xi} u(\xi) {\xi+t \over \xi-t} = E(t-t^{-1})
\label{sixteen} 
\end{equation}
Equation (\ref{sixteen}) specifies the Reimann-Hilbert problem for the
arc $(t_1,t_2)$: to find an analytic function in the cut plane whose
boundary values satisfy (\ref{sixteen}).  In our case we have an
additional condition to satisfy at $z = \infty$ which follows from the
normalization of $u(t)$, 
\begin{equation}
\Phi(\infty) = -1
\label{seventeen}
\end{equation}
The discontinuity formula (\ref{fifteen}) (a reflection of Gauss' law
in electro-statics) implies that a solution to the Reimann-Hilbert
problem solves the integral equation (\ref{thirteen}).

The function $h(z) = [(z-t_1)(z-t_2)]^{1/2}$ clearly solves the
homogeneous problem $\phi^+(t) + \phi^-(t) = 0$ up to an entire
function.  This choice of the square root function vanishing at the
end points leads to $u(t_1) = u(t_2) = 0$.  Now write 
\[
\Phi(z) = h(z) H(z) 
\]
substitution in (\ref{sixteen}) gives
\begin{equation}
H^+(t) - H^-(t) = {E(t-t^{-1}) \over \sqrt{(t-t_1)(t-t_2)}}
\label{eighteen}
\end{equation}
(By the square root in the discontinuity formula for $H(z)$ we will
mean $h^+(t) = \exp i\theta/2\left[\sin^2 \d{\alpha \over 2} - \sin^2
\d{\theta \over 2}\right]^{1/2}$.)  From (\ref{seventeen}) we know
that $H(\infty) = 0$.  Hence $H(z)$ is uniquely given by the Plemelj
formulae 
\[
H(z) = {1 \over 2\pi i} \int^{t_2}_{t_1} {E(t-t^{-1}) \over
\sqrt{(t-t_1) (t-t_2)}} {dt \over t-z}
\]
This integral can be evaluated by standard methods.  Note
\[
H(z) = {1 \over 4\pi i} \int_\Gamma {E(t-t^{-1}) \over \sqrt{(t-t_1)
(t-t_2)}} {dt \over t-z}
\]
where $\Gamma$ encloses the cut $(t_1,t_2)$ and $z$ lies outside it.
This is always possible since $z \ {\epsilon \!\! /} \ (t_1,t_2)$.
$H(z)$ has poles at $t=0$ and $t=z$ with residues $E/2z$ and
$E(z-z^{-1})/2[(z-t_1) (z-t_2)]^{1/2}$ respectively.  The contribution
from infinity is $E/2$ where we have taken the branch
\[
\sqrt{(z-t_1)(z-t_2)} \rightarrow -z \ \ \ {\rm as} \ z \rightarrow
\infty 
\]
Hence
\[
H(z) = {E \over 2} \left[{z-z^{-1} \over \sqrt{(z-t_1)(z-t_2)}} + 1 +
{1 \over 2}\right]
\]
and
\[
\Phi(z) = {E \over 2} \left(z + {1 \over z}\right) + {E \over 2}
\left({1 \over z} + 1\right) \sqrt{(z-t_1)(z-t_2)}
\]
$\Phi(\infty) = -1$ implies 
\[
{1 - \cos \alpha \over 2} = {1 \over E} = \sin^2 {\alpha \over 2},
\ \ \ 1 \leq E < \infty
\]
From the discontinuity formula (\ref{fifteen}) we get 
\[
u(t) = {E \over 2} \left({1 \over t} + 1\right)h^+ (t) 
\]
\[
u(\theta) = 2E \cos {\theta \over 2} \sqrt{E^{-1} - \sin^2 {\theta
\over 2}} \geq 0
\]
We summarize the computation 
\[\hspace*{.5cm}
u(\theta) = \left\{\matrix{1 + E \cos \theta \hspace*{2.5cm}
& \hspace*{1cm} 0 \leq E \leq 1 & \hspace*{2cm} (19a) 
\cr \cr
2E \cos \d{\theta \over 2} \sqrt{E^{-1} - \sin^2 \theta/2}
& \hspace*{1cm} 1 \leq E < \infty & \hspace*{2cm} (19b)}\right.
\]
Our expectation for the appearance of a gap at $\theta = \pi$ beyond
$E = 1$ is confirmed; also the size of the gap $2\alpha$ is a
continuous and monotonic function of $E$: $1/E = \sin^2 \alpha/2$.
\bigskip

\noindent {\Large{\bf IV. Computation of Thermodynamic Quantities in
the} $N \rightarrow \infty$ {\bf Limit}}
\bigskip

The appearance of the gap at $E = 1$ is reflected in the behavior of
the thermodynamic functions.  From (\ref{seven}) and (\ref{nine}) we
have for the free energy per degree of freedom 
\[\hspace*{1cm}
f(E) = \left\{\matrix{\d{E^2 \over 4} \hspace*{2cm} & \hspace*{1.5cm} 0
\leq E \leq 1 & \hspace*{2.5cm} (20a) \cr \cr E - \d{1\over2} \log E -
\d{3\over4} & \hspace*{1.7cm} 1 \leq E < \infty & \hspace*{2.5cm}
(20b)} \right. 
\]
$f(E)$ and its first two derivatives are continuous at $E=1$,
indicating a continuous phase transition at the point $E = 1$.

A useful order parameter$^{9}$ that characterizes the phases is   
$$
R = \sum^N_{i=1} {1 \over N} \sin^2 \theta_i = {1 \over N} {\rm Tr}
\left[{U - U^+ \over 2i}\right]^2
\eqno (21)
$$
In the continuum approximation as $N \rightarrow \infty$ we have
$$
R(E) = \lim_{N \rightarrow \infty} \langle {1 \over N} \sum^N_{i=1} \sin^2
\theta_i \rangle = \int^{2\pi}_0 {d\theta \over 2\pi} u(\theta) \sin^2
\theta 
\eqno (22)
$$
Therefore
$$
R(E) = \left\{\matrix{1/2 \hspace*{1.1cm} & 0 \leq E \leq 1 \cr \cr
\d{1 \over E} - \d{1 \over 2E^2} & 1 \leq E < \infty}\right.
\eqno (23)
$$
$R$ represents the average value of the sum of the squares of the
$y$-co-ordinates of the charges.  $R(E)$ and $R'(E)$ are continuous at
$E = 1$.  In a sense, $R$ is a disorder parameter because it is a
measure of the randomness of the system.  In the phase $0 \leq E \leq
1$ where its expectation value is a constant, the distribution of
eigenvalues of the unitary matrix is truly random and the eigenvalues
go over their entire range.  We shall call this phase the random phase
(RP).  In the phase $1 \leq E < \infty$ we see that for $E \gg 1$,
$\langle R \rangle \approx 0$; the random distribution has
continuously disappeared and the eigenvalues are confined to a small
region around $U=1$.  Also the eigenvalue density (196) for $E \gg 1$
is written in scaled form
$$
u(\theta) \approx 2\sqrt{E} \sqrt{1 - {(\sqrt{E}\theta)^2 \over 4}},
\ \ \ |\theta| \leq {2 \over \sqrt{E}}
\eqno (24)
$$
We call this phase the familiar spin wave phase (SWP).  We are aware
that for $E \gapprox 1$, the term SWP is certainly not very
appropriate. 
\bigskip

\noindent {\Large{\bf V. The Wilson Loop}}
\bigskip

An important order parameter of the original $U(N)$ theory is the
Wilson loop for a closed curve $C$, 
$$
W[C] = {1 \over ZN} \int d\mu {\rm Tr}\left(\prod_{\ell}
U_\ell\right) e^{\beta({\rm Tr} U + {\rm Tr} U^+-N)}
\eqno (25)
$$
Once more, because in two dimensions plaquette variables are the
independent variables, we have for a simple curve 
\[
W(C) = \omega^A
\]
$$
\omega = {1 \over ZN} \int dU({\rm Tr} U) e^{\beta({\rm Tr} U + {\rm
Tr} U^+)}
\eqno (26)
$$
$A$ is the area enclosed by the curve in lattice units.  In the
diagonal representation 
$$
\omega = \left\langle {1 \over N} \sum^N_{i=1} \cos \theta_i
\right\rangle 
\eqno (27)
$$
which in the continuum approximation of the large $N$ limit becomes
\[
= \lim_{N \rightarrow \infty} \left\langle {1 \over N} \sum^N_{i=1}
\cos \theta_i \right\rangle = \int^{2\pi}_0 {d\theta \over 2\pi}
u(\theta) \cos \theta.
\]
Using (19) we have 
$$
W_A = \left\{\matrix{\left({E \over 2}\right)^A \hspace*{1.1cm}
& \hspace*{1cm} 0 \leq E \leq 1 \cr
\cr \left(1 - {1 \over 2E}\right)^A & \hspace*{1cm} 1 \leq E < \infty}\right.
\eqno (28)
$$ 
The string constant in lattice units is defined by 
\[
W_A = e^{{-A} {1 \over 4\pi\alpha'_0}}
\]
$$
\alpha_0 = \left\{\matrix{-\d{1 \over 4\pi} \d{1 \over \log
E/2} \hspace*{1.2cm} & \hspace*{1cm} 0 \leq E \leq 1 
\cr \cr -\d{1 \over 4\pi} \d{1 \over \log\left(1 - \d{1
\over 2E}\right)} & \hspace*{1cm} 1 \leq E < \infty}\right.
\eqno (29)
$$
it is the exact analogue of the correlation `length' in spin systems.
Up until now we have worked from the statistical mechanics point of
view in which the basic unit of length is the lattice spacing `$a$'.
In order to make contact with field theory$^{10}$ where the basic unit
of length is usually set by the inverse renormalized mass we go over
to laboratory units in which the string constant has the dimension of
area 
$$
\alpha(a;E) = a^2 \alpha_0(E)
\eqno (30)
$$
This physical quantity is a renormalization group invariant i.e., if
$a \rightarrow \lambda a$, $\lambda > 0$
$$
\alpha (\lambda a; E) = \alpha(a; R_\lambda (E))
\eqno (31)
$$
$R_\lambda (E)$ is the renormalized coupling.  Then from (29) we have
$$
R_\lambda (E) = \left\{\matrix{2\left(\d{E \over 2}\right)^{1/\lambda^2} 
\hspace*{2cm} & \hspace*{1cm} 0 \leq E \leq 1 \cr \cr \d{1 \over 2\left[1
- \left(1 - \d{1 \over 2E}\right)^{1/\lambda^2}\right]}
& \hspace*{1cm} 1 \leq E < \infty}\right. 
\eqno (32)
$$
Now consider moving towards short distances, e.g., $\lambda =
\d{1\over2} < 1$.  The recursion formulae which follows from (32) have
two trivial fixed points $E^\star = 0$ and $E^\star = \infty$ and the
flow is away from $E^\star = 0$ towards $E^\star = \infty$.  Note that
$E = 1$ is \underbar{not} a fixed point.  At the fixed points we have 
\begin{eqnarray}
\alpha_0 (E^\star = 0) &=& 0 \nonumber \\[2mm] 
\alpha_0(E^\star = \infty) &=& \infty \nonumber
\end{eqnarray}
Hence $E^\star = \infty$ is the critical point of the statistical
mechanics problem for fixed $a$.  In the neighborhood of the critical
point $E^\star = \infty$, the lattice constant is an irrelevant
parameter and it is possible to define a euclidean invariant theory
characterized by the adjustable parameter
\[
\alpha_R = \alpha (a \rightarrow 0, E \rightarrow E^\star)
\]
From (29) we have for $\alpha_R$ near the critical point
\[
\alpha_R = {Ea^2 \over 2\pi} \equiv {E_R \over 2\pi}
\]
In the field theory notation $\d{E \over 2} = \d{\beta \over N} = \d{1
\over g^2_0 N}$, hence 
\[
\alpha_R = {1 \over \pi N(g^2_0/a^2)}
\]
The renormalized parameter $\alpha_R$ ranges from 0 to $\infty$, in
contrast to the bare parameter $\alpha_0$ which ranges from $1/2\pi$
to $\infty$.  This agrees with `t Hooft's computation$^2$ of the same
quantity in the Yang-Mills version of the theory.  In fact, for the
interaction energy of static quarks in the fundamental representation,
we have 
\[
E(x_1,x_2) = {1 \over 4\pi\alpha_R} |x_1 - x_2|
\]
$x_1$ and $x_2$ are the locations of the quarks in one dimension. 
\bigskip\bigskip
\newpage

\noindent {\Large{\bf VI. Partition Function Near Critical Point}} 
\bigskip

The partition function $z$ in (\ref{six}) can be written by expanding
the cosine and the log in their respective power series.  The action
is 
\begin{eqnarray} 
-2S &=& E \sum^N_{i=1} {\cos \theta_i \over N} + \sum_{i<j} {1 \over
N^2} \log 2[1 - \cos(\theta_i - \theta_i)] \nonumber \\[2mm] 
&=& E \sum^N_{i=1} {1 \over N}\left[1 - {\theta^2_i \over 2!} +
{\theta^4_i \over 4!} \cdots\right] + \nonumber \\[2mm] 
&& \sum_{i<j} {1 \over N^2} \log(\theta_i - \theta_j)^2 - \sum_{i<j}
{1 \over N^2} {(\theta_i - \theta_j)^2 \over 3.4} \nonumber
\end{eqnarray}
Now performing a scale transformation
\[
\theta_i \rightarrow \sqrt{E} \theta_i = \xi_i
\]
we get
\begin{eqnarray}
-2S &=& - {1 \over N} \sum^N_{i=1} \xi^2_i/2 + {1 \over N^2}
\sum_{i<j} \log|\xi_i - \xi_j|^2 \nonumber \\[2mm]
&& + E - {N(N-1) \over 2N^2} \log E + \sum^\infty_{n=1} \left({1
\over E}\right)^{\theta(n)} [\xi_i] \nonumber
\end{eqnarray}
$\theta^{(n)}$ stands for the coefficient of $1/E^n$, e.g.,
\begin{eqnarray}
\theta^{(1)} &=& {1 \over N} \sum_i \xi^4_i - {1 \over N^2} \sum_{i<j}
{(\xi_i - \xi_j)^2 \over 12} \nonumber \\[2mm] 
\theta^{(2)} &=& - {1 \over N} \sum_i \xi^6_i - {1 \over N^2}
\sum_{i<j} {(\xi_i - \xi_j)^4 \over 3.4.5.6} \cdots {\rm etc.}
\nonumber
\end{eqnarray}
After scaling the measure factor we have 
\begin{eqnarray}
z &=& e^{N^2\left(E - {1\over 2}\log E\right)}{1 \over N!}
\int^{+\infty}_{-\infty} \prod_i {d\xi_i \over 2\pi} \prod_i
\theta(\sqrt{E} \pi - |\xi_i|) \nonumber \\[2mm] 
&& \times e^{N^2V} \cdot e^{N^2 \sum^\infty_{n=1}} {\theta^{(n)} \over E^n}
(\xi) \nonumber
\end{eqnarray}
$$
V = -{1 \over N} \sum^N_{i=1} {\xi^2_i \over 2} + {1 \over N^2}
\sum_{i<j} \log |\xi_i - \xi_j|^2 
\eqno (33)
$$
The $\theta$ function in (33) is a cut-off that reflects the
compactness of the gauge group.

So far, (33) is exact without approximation.  In the large $N$ limit,
one can once more consider a WKB expansion in $1/N^2$.  Since this is
the same problem as before except for a change of scale, we have, for
the distribution of eigenvalues,
$$
{\tilde u}(\xi) = {1 \over \sqrt{E}} u(\xi/\sqrt{E}) = 2 \cos {\xi
\over 2\sqrt{E}} \sqrt{1 - E \sin^2 \xi/2 \sqrt{E}}
\eqno (34)
$$
The function $u$ above is given by (19b), since we are, ultimately,
interested in large values of $E$.  Expanding, we have
\[
{\tilde u}(\xi) = \sqrt{4 - \xi^2} + \sum^\infty_{n=1} {C_n (\xi)
\over E^n}, \ \ \ |\xi| \leq 2, \ E \gg 1
\]
This means that the contribution of the terms involving the operators
$\theta^{(n)} (\xi_i)$ in (33), to the leading term in the WKB
expansion in $1/N^2$, are negligible near critical point i.e., as $E
\rightarrow \infty$ (N.B. the large $E$ limit is taken
\underbar{after} the large $N$ limit).  Further, the cut-off in (33)
is automatically repected, and we have 
$$
z (E \rightarrow E^\star) \approx e^{N^2\left(E - {1\over 2} \log
E\right)} {1 \over N!} \int^\infty_{-\infty} \prod_i {d\xi_i \over
2\pi} e^{N^2V}  
\eqno (35)
$$

We now recall that if $H$ is a $N \times N$ hermetian matrix with
$N^2$ independent components, the integration measure over $H$ is
defined by$^7$ 
$$
dH = \prod^N_{i=1} dH_{ii} \prod_{i<j} dH^R_{ij} dH^I_{ij},
\ \ \ H_{ij} = H^R_{ij} + i H^I_{ij} 
\eqno (36)
$$
Once more one can perform a separation of variables to extract those
which are left invariant by a unitary transformation.  Let $W$ be the
unitary matrix that diagonalizes $H$,
\[
H = W^+h \ W, \ \ \ \ \ h_{ij} = \delta_{ij} \xi_j
\]
then
\[
dH \propto dW. \ \ \ \prod^N_{i=1} d\xi_i \prod_{i<j} |\xi_i -
\xi_j|^2 
\]
and (35) becomes
$$
z \propto e^{N^2(E - \log E/2)} \int dH e^{-{N \over 2} {\rm Tr}
H^2} 
\eqno (37)
$$
Therefore, the partition function of the original gauge theory near
critical point is
\begin{eqnarray}
Z &=& e^{-2N\beta V} z^V \nonumber \\[2mm] 
&\propto& e^{-N^2 V \log E/2} \left[\int dH e^{-{N \over 2} {\rm Tr}
H^2}\right]^V \nonumber
\end{eqnarray}
$$
\propto \ \left[ \int {dH \over (\sqrt{E})^{N^2}} e^{-{N \over 2} {\rm Tr}
H^2}\right]^V \hspace*{.7cm}
\eqno (38)
$$
Expression (38) is what one expects to be proportional to the
partition function of the Yang-Mills version of the theory, which is
defined in a spacetime volume $V$ by 
$$
Z_{YM} \propto \int \prod_{x,t} dA_0 dA_1 \delta(A_0(x,t))
\delta(A_1(x,t_0)) e^{-{{\rm Tr} \over 4g^2} \d\int_V dx dt
F^2_{\mu\nu}}  
\eqno (39)
$$
${\vec F}_{\mu\nu}$ is the field strength; $g$ is the dimensional
coupling constant.  We have fixed the generalized axial gauge 
\begin{eqnarray}
A_0(x,t) &=& 0 \nonumber\\[2mm] 
A_1(x,t_0) &=& 0 \nonumber
\end{eqnarray}
In this gauge one can express $A_1$ uniquely in terms of the field
strength 
$$
A_1(x,t) = \int^t_{t_0} F_{01}(x,t') dt'
\eqno (40)
$$
Making this change of variable in (39)
\[
Z_{YM} \propto \int \prod_{x,t} dF(x,t) {e}^{-\frac{1}{ 2g^2}
\int_V dx dt {\rm Tr} F^2(x,t)} 
\]
At this stage we can manipulate using a cut-off to write 
\[
Z_{YM} \propto \left(\int dF e^{-\frac{a^2}{ 2g^2}} {\rm Tr}
F^2\right)^V 
\]
$dF$ is the integration measure of the Hermetian matrix $F$.
Introducing the dimensionless coupling $g^2_0 = g^2 a^2$, which
vanishes as $a \rightarrow 0$, and the dimensionless matrix $H = 2a^2
F/\sqrt{E}$; $E = 2/g^2_0 N$ a fixed number as $g_0 \rightarrow 0$ and
$N \rightarrow \infty$, we have
$$
Z_{YM} \propto \left(\int {dH \over (\sqrt{E})^{N^2}} e^{-{N\over 2}{\rm
Tr} H^2}\right)^V
\eqno (41)
$$
Hence the formal equivalence of (38) and (39) is established.

Using entirely similar arguments it is easy to show that in the
thermodynamic limit of large $N$, near the critical point, the Wilson
loop is given by 
$$
W[C] = \left[1 - {1 \over 2E} \left\langle {1 \over N} \sum^N_{i=1}
\xi^2_i \right \rangle\right]^A
\eqno (42)
$$ 
$$
\left\langle {1 \over N} \sum^N_{i=1} \xi^2_1\right\rangle = {\d\int
\d\prod_i {d\xi_i \over 2\pi} \d\prod_{i<j} |\xi_i - \xi_i|^2 \d\sum^N_{i=1}
\xi^2_1/N  e^{-{N^2 \over 2} \d\sum_i \xi^2_1/N} \over \d\int
\d\prod_i {d\xi_i \over 2} \d\prod_{i<j} |\xi_i - \xi_i|^2 e^{-{N^2
\over 2} \d\sum_i \xi^2_i/N}}
\eqno (43)
$$
(N.B. to leading order in WKB $\left\langle \d{1\over N} \d{\sum_i}
\xi^2_1\right\rangle = 1$, in agreement with 28).  (42) can be
rewritten as an integral over a random Hermetian matrix $H$: 
\[
W[C] = \left[1 - {1 \over 2E} \left\langle {1 \over N} {\rm Tr}
H^2\right\rangle\right]^A
\]
$$
\eqno (44)
$$
\[
\left\langle {1 \over N} {\rm Tr} H^2\right\rangle = 
{{\d\int
dH} {\left({1\over N} {\rm Tr} H^2\right)e^-{N\over 2}{\rm Tr}
H^2}} \over {\d\int
dH} e^-{N\over 2}{\rm Tr} H^2
\]
(44) is exactly the expression for the Wilson loop in the Yang-Mills
theory.  The steps are similar to those which led from (39) to (41). 

\newpage

\noindent {\Large{\bf VII. Exact Computation of Partition Function for
Fixed $\beta$ and $N$}}
\bigskip

So far, we have studied the $U(N)$ gauge theory in the limit $N
\rightarrow \infty$, holding the ratio $E = 2\beta/N$ fixed.  We now
wish to compare these results with computations for finite $N$.

The partition function $z$ can be expressed as a Toeplitz determinant.
The trick lies in expressing the repulsive measure factor in (4) as a
product of Vandermonde determinants 
\[
\prod_{i<j} (e^{i\theta_i} - e^{i\theta_j}) \prod_{i<j}
(e^{-i\theta_i} - e^{-i\theta_j}) = \det V \det V^\star
\]
$$
V_{\ell m} = e^{im\theta_\ell} \hspace*{5cm}  
\eqno (45)
$$
\[
\det V = \sum_{J_1 \cdots J_n} \epsilon_{J_1 \cdots J_N} V_{1J_1
\cdots V_{NJ_N}}
\]
Substituting (45) in (4) we have
\[
z = {1 \over N!} \sum_{J_1 \cdots J_N} \epsilon_{J_1 \cdots J_N}
\epsilon_{K_1 \cdots K_N} \int^{2\pi}_0 \prod^N_{i=1} {d\theta_i \over
2\pi} e^{i\d\sum_\ell \left[(j_\ell - k_\ell)\theta_\ell + 2\beta
\cos \theta_\ell\right]}
\]
$$
= \det M \hspace*{9cm}
\eqno (46)
$$
\[
M_{ij} = I_{i-j} (2\beta) = \int^{2\pi}_0 {d\theta \over 2\pi}
e^{i(k-j)\theta + 2\beta\cos \theta} \hspace*{3.7cm}
\]
$I_n (2\beta)$ is a Bessel function with imaginary argument$^{11}$.

It is clear that as long as $N$ is finite, $z$ is an analytic function
of $\beta$.  It is also known that the $U(N)$ gauge theory with finite
$N$ has only 2 trivial fixed points of the renormalization group$^6$,
$\beta = 0$ and $\beta = \infty$.  As one goes to short distances, the
coupling constant renormalizes to larger values and $\beta = \infty$
is a critical point near which one can define a continuum theory. 

From formula (26) we have, for the Wilson loop,
\[
W[C] = \omega^A \hspace*{1cm}
\]
$$
\omega(\beta,N) = {1 \over 2N} {\partial \over \partial \beta} \log
z 
\eqno (47)
$$
It is instructive to compute $\omega(\beta,N)$ near the fixed points
$\beta = 0$ and $\beta = \infty$. 

A Taylor series expansion of the function
$$
F_N(\beta) = z(\beta,N) = \det(I_{i-j}(2\beta))_{N\times N}
\eqno (48)
$$
was done on a computer around the fixed points $\beta = 0$ and $\beta
= \infty$ for $N = 2,3,4,5$.  This calculation was done by Mark
Sweeny.  The result are 
\bigskip

\noindent (i) Taylor series around $\beta = 0$:
\[
F_2(\beta) = 1 + \beta^2 + {\beta^4 \over 2} + {5 \over 36} \beta^6 +
{7 \over 288} \beta^8 + \cdots
\]
$$
F_3 (\beta) = 1 + \beta^2 + {\beta^4 \over 2} + {\beta^6 \over 6} +
{23 \over 576} \beta^8 + \cdots
\eqno (49)
$$
\[
F_4(\beta) = 1 + \beta^2 + {\beta^4 \over 2} + {\beta^6 \over 6} + {1
\over 24} \beta^4 + {119 \over 14400} \beta^{10} + \cdots 
\]
\[
F_5(\beta) = 1 + \beta^2 + {\beta^4 \over 2} + {\beta^6 \over 6} + {1
\over 24} \beta^8 + {1 \over 120} \beta^{10} + \cdots
\]
We see that
$$
F_N (\beta) \sim e^{\beta^2} 
\eqno (50)
$$
is a fairly accurate fit that grows better with increase in $N$.  From
(50) we can compute 
$$
\omega(\beta,N) \simeq {\beta \over N} = {E \over 2} 
\eqno (51)
$$

\noindent (ii) Taylor series around $\beta = \infty$:

Defining 
\[
F_N(\infty) = \left({e^{2\beta} \over \sqrt{4\pi \beta}}\right)^N
f_N(x), \ \ \ x = \beta^{-1}
\]
The Taylor expansion of $f_N(x)$ around $x = 0$ is
\[
f_x(x) = 4x - 7x^2 + {61 \over 8} x^3 - {151 \over 32} x^4 + {771
\over 512} x^5 \cdots 
\]
$$
f_3 (x) = 128 x^3 - 912 x^4 + 3297 x^5
\eqno (52)
$$
\[
f_4 (x) = 49152 x^6 - 860160 x^7 + 7452672 x^8 + \cdots 
\]
Note that the powers of the leading terms are given by $N(N-1)/2$.
This leads to
\[
\omega (\beta,N) \simeq 1 - {1 \over 2} {N \over 2\beta} + O\left({1
\over \beta^2}\right)
\]
$$
\hspace*{1cm} \cong 1 - {1 \over 2E}, \ \ \ \beta \gg 1
\eqno (53)
$$
The small corrections are computable from (52).  We note that this
result is also inferred from the exponential factor that scales $z$ in
(33).  We do not know a simple explanation for the linear fit in (51).

We see that the finite $N$ results (51) and (53) near the fixed points
exactly match the corresponding expressions (28) computed in the
thermodynamic limit of $N \rightarrow \infty$.
\bigskip

\noindent {\Large{\bf VIII. Conclusion and Discussion}}
\bigskip

The analytic computation done in the large $N$ limit holding $E =
2\beta/N$ fixed is in very good agreement with the computation for
finite $N$, especially near the fixed points and it is important to
emphasize this agreement in the critical region which is the
physically relevant one.  The difference is that while the computation
for finite $N$ involves analytic interpolation between weak and strong
coupling the thermodynamic computation shows non-analytic behavior at
$E = 1$.  Certainly this non-analyticity is a pathology of the $N =
\infty$ limit.  We do not know whether the qualitative agreement of
the two methods and even the anomaly at $E = 1$, is a feature of the
theory in higher dimensions which is vastly more complicated.  It is
desirable to express the summation of the planar diagrams in higher
dimensions as a coupled set of integral equations in the mean field
approximation. 

We end with a final comment that the expression of the partition
function as a determinant implies that this theory has a fermionic
representation. 
\bigskip

\noindent {\Large{\bf Acknowledgement}}
\bigskip

I wish to thank Professor Yoichiro Nambu and Professor Leo Kadanoff
for useful discussions and suggestions.  
I have also had some useful discussions with
Nino Brali\'c, Tohru Eguchi, Mark Sweeny, Jorge Jos\'e and Don
Weingarten.


\newpage

\begin{thebibliography}{999}
\bibitem{one}
See, for example: a) W.A. Bardeen, B.W. Lee and R.E. Schrock,
Phys. Rev. \underbar{D14}, 985 (1976); b) E. Br\'ezin, J.C. Le Guillou
and J. Zinn-Justin in ``Phase Transitions and Critical Phenomena'',
Vol. 6, Edited by C. Domb and M.S. Green; c) S.K. Ma's article in the
same volume.
\bibitem{two}
G. `t Hooft, Nucl. Phys. \underbar{B72}, 461 (1974). \\ G. `t Hooft,
Nucl. Phys. \underbar{B75}, 461 (1975).
\bibitem{three}
E. Witten Havard Preprints HUTP-79/A007 and HUTP/79/A014.
\bibitem{four}
E. Br\'ezin, C. Itzykson, G. Parisi and J.B. Zuber,
Commun. Math. Phys. \underbar{59}, 35 (1978).
\bibitem{six}
K.G. Wilson, Phys. Rev. \underbar{D10}, 2245 (1974); Erice Lecture
notes 1975. 
\bibitem{seven} 
J.M. Drouffe and C. Itzykson, Physics Reports \underbar{38C}, No. 3
(1978). 
\bibitem{eight}
Integration over matrices has been extensively discussed in Nuclear
Physics.  See, e.g., F.J. Dyson, J. Math. Phys. \underbar{3}, (1962)
reprinted in `Statistical Theories of Spectra: Fluctuations' C. Porter
(Academic Press, 1965); `Random Matrices', M.L. Mehta (Academic Press,
1967). 
\bibitem{nine}
`Singular Integral Equations', N.I. Muskhelishvili (P. Noordhoof
N.V. - Groningen - Holland). 
\bibitem{ten}
Suggested by L. Kadanoff.
\bibitem{elevan}
See, e.g., K.G. Wilson and J. Kogut, Phys. Reports \underbar{12C}
(1974).  L.P. Kadanoff, Rev. Mod. Phys. \underbar{49}, (1977).
 \bibitem{twelve}
After this work was well begun we received a preprint on $U(N)$
lattice gauge theories by I. Bars and F. Green (IAS preprint) which
also contains formula (40).  In our notation they have computed the
determinant for small $E = \d{2\beta \over N}$.  We remark that in
this limit the determinant can be exactly computed using the classical
theorem of Szeg\"o on Toeplitz determinant:  For large $N$, fixed
$\beta$ 
\begin{eqnarray}
&\log& \!\!\!\!\! ~{\rm det}~ I_{i-j} (2\beta) \sim Ng_0 +
\d{\sum^\infty_{n=1}} ng_n g_{-n} = \beta^2 \nonumber \\[2mm] 
&g_n& \!\!\! = \int^{2\pi}_0 {d\theta \over 2\pi} e^{in\theta} \log
(e^{2\beta\cos\theta}) \nonumber
\end{eqnarray}
(See McCoy and Wu, `The Two Dimensional Ising Model [Hardvard
Univ. Press], for an extensive discussion). \\ Other recent papers
that discuss $U(N)$ lattice theories: T. Eguchi (Chicago preprint EFI
79/22); D. Weingarten (Indiana preprint).
\bibitem{thirteen} 
I thank Mark Sweeny for responding with interest to do the computer
calculation. 
\bibitem{five}
While this work was going on and after the solutions of the integral equations were obtained we learnt of a similar work by D. J. Gross and E. Witten (Princeton preprint)
\end{thebibliography}
\end{document}